\documentclass[aps,pra,fleqn,showpacs,showkeys,twocolumn]{revtex4-1} %,preprint
\usepackage[pdftex]{graphicx}
\usepackage{amssymb}
\usepackage{amsmath}
\usepackage{bm}
\begin{document}
\setcounter{page}{1}
\title[]{Photon Emission Dynamics of a Two-Level Atom in a Cavity}
\author{Chang Jae \surname{Lee}}
\email{coolcjl@sunmoon.ac.kr}
\thanks{ Tel:+82-041-530-2243}
\affiliation{Department of Nanochemistry, Sunmoon University, Asan 336-708}
\date[]{}
\begin{abstract}
The collapse and revival of quantum states appear in diverse areas of physics. In quantum optics the occurrence of such a phenomena in the evolution of an atomic state, interacting with a light field initially in a coherent state, was predicted by using the Jaynes-Cummings model (JCM), and subsequently demonstrated experimentally. In this paper we revisit the JCM with the Monte-Carlo wave function approach and investigate the time evolution of the photon emission rate of  the atom in a cavity. Analytical and numerical quantum trajectory calculations show that the cavity and the initial field statistics strongly influence the photon emission dynamics. A coherent field indeed gives rise to a collapse and revival behavior that mirrors atomic state evolution. However, there are differences between the two. The emission rate for a field in a Fock number state exhibits a sinusoidal oscillation, and there exists a quiescent period for a thermal field. These properties are quite different from those in free space. It is also shown that the fluctuation in photon emission is much less than that of the atomic population.
\end{abstract}
\maketitle
\section{INTRODUCTION}

The collapse and revival of quantum states by either forced or self-regulated processes is a fascinating phenomena with a long history. Hahn discovered spin echo~\cite{spin_echo} in 1950, and photon echoes were detected at optical frequencies in mid 1960s~\cite{photon_echo}. Analogous echo phenomena can be found in vibrational spectroscopy~\cite{vib_echo} and in the dynamics of an atomic matter wave in a light field~\cite{atom_echo, tppi}.  Echo techniques are of great interest in quantum computing and quantum state engineering applications where maintaining coherence is a critical requirement~\cite{decoherence}. The time evolution of the population inversion of a two-level atom (TLA) interacting with a single-mode light field, as modeled by the JC Hamiltonian~\cite{jcm}, also shows the collapse~\cite{cum} and revival behavior when the atom interacts with a field that is  initially in a coherent state~\cite{eberly, rempe}.  The JCM, despite simplicity and age, still provides a fertile ground for testing foundations of quantum theory and prototyping practical applications~\cite{jopb_jcm}.

Previously, we approached the JCM with a viewpoint based on the Monte-Carlo Wave Function (MCWF) method~\cite{mcwf}, and found that a TLA interacting with a field initially in a number state emits (and absorbs) photons with a unique counting statistics~\cite{jkps}. In this paper, we revisit the JCM with fields in coherent and thermal states in addition to the number state.  The goal is to examine what effects the field statistics has on emitted photons, rather than the usual atomic state dynamics. In the next section, we briefly describe our adaptation of the MCWF method, and after that, results of MCWF simulations are given along with a quantum trajectory analysis on the results. Finally, the main discoveries and further discussions are given in Conclusions section.

\section{MCWF Method for Photon Emission}
\subsection{MCWF approach for the JCM dynamics}
We consider a fully-quantized Hamiltonian for a system of a TLA interacting with a single mode field in a lossless cavity.
The Hamiltonian considers only the internal energy, neglecting the center-of-mass motion of the atom. With the use of the rotating-wave approximation for the interaction between the TLA and the quantized-field, the total Hamiltonian is given by:
\begin{equation}
H = \frac{1}{2} \hbar\omega_{0}\sigma_{z} + \hbar\omega a^{\dagger}a +  \frac{\hbar\Omega_{0}}{2} (a \sigma_{+} + a^{\dagger} \sigma_{-}).
\label{eq:jcm}
\end{equation}
In the above, $\omega_{0} = (E_{e} - E_{g})/\hbar$ is the frequency between the upper and the lower states of the atom, $|e \rangle$ and $|g \rangle$. The frequency of the field is $\omega$, and $a$ and $a^{\dagger}$ are the field annihilation and creation operators, respectively, with the Fock number state being the eigenstate of the photon number operator, $a^{\dagger} a | n \rangle = n | n \rangle$. The Pauli matrices $\sigma_{z}, \sigma_{\pm}$ are operators for the atomic population and transitions
$$
\sigma_{z} = \begin{pmatrix}
1 & 0 \\
0 & -1
\end{pmatrix} ,
\sigma_{+} = \begin{pmatrix}
0 & 1\\
0 & 0
\end{pmatrix},
\sigma_{-} = \begin{pmatrix}
0 & 0\\
1 & 0
\end{pmatrix},
$$
and $\Omega_{0}$ is the vacuum Rabi frequency for the atom-field interaction.

The evolution of the TLA-field system is governed by
\begin{equation}
| \psi (t) \rangle = \exp \left( - \frac{i }{\hbar} H t \right) | \psi (0) \rangle .
\label{eq:evol}
\end{equation}
In case that the atom is initially in the upper state and the field in the number state  $|n \rangle $, namely $ | \psi (0) \rangle =  |n , e \rangle $, the transition probability of the system to $| n+1 , g \rangle$ is given by~\cite{louisell}
\begin{equation}
P_{g}(t) = {\left| \langle n+1 , g | \psi(t) \rangle \right|}^{2}=\frac{{\Omega_{n}}^{2}}{ {\Omega_{\rm eff}}^{2}}\  \sin^{2}\, \frac{\Omega_{\rm eff} }{2}t,
\label{eq:transProb}
\end{equation}
where $\Omega_{n} = \Omega_{0}\ \sqrt{n+1}$ is the $n-$photon Rabi frequency, $\Omega_{\rm eff}={\sqrt{(\Delta\omega )^{2} + {\Omega_{n}}^{2} }}$, with $\Delta \omega = \omega_{0} - \omega$ being the detuning. We are not interested in the effects of detuning in this paper, so we will set it to zero: $\Delta \omega  =0$. In that case Eq. (\ref{eq:transProb}) simplifies to
\begin{equation}
P_{g}(t) = \sin^{2}\, \frac{ \Omega_{n}}{2} t,
\label{eq:tpsimple}
\end{equation}
and the probability of transition to the upper state is $P_{e}(t) = 1-P_{g}(t)$.

For a general initial field state with a photon number distribution $p_{n}(0)$, the probability of transition to the upper state at resonance becomes
\begin{equation}
P_{e}(t) = \frac{1}{2} \left[ 1 + \sum_{n=0}^{\infty} p_{n}(0)\ \cos \, \Omega_{n} t \right].
\label{eq:pe}
\end{equation}

Our MCWF approach is described in detail in Refs.~\cite{bkcs,jkps}, so we will give only an outline here.
In the MCWF approach, the system is in the superposition state, Eq.\eqref{eq:evol},  before a measurement is made, and the measurement causes the state jump to either $ |n, e \rangle$ or $|n+1, g \rangle$.
The algorithm to simulate the reduction of the state may be summarized as follows:
\begin{enumerate}
\item Divide the interaction time (the Rabi cycle) into $N$ segments: $N\Delta t = 2 \pi$.
\item At time $t_{m} = t_{0} + m \Delta t, \   (t_{0}=0; m = 1, 2, \ldots , N)$ generate a random number $r_{m}$ in the range $[0,1]$.
\item Compute $P_{e}(t_{m})$ given by Eq.~(\ref{eq:pe}) and compare it with $r_{m}$.
\begin{enumerate}
\item If $P_{e}(t_{m}) > r_{m}$, the quantum jump $| \psi (t) \rightarrow |n, e \rangle$ occurs.
\item If $P_{e}(t_{m}) < r_{m}$, the quantum jump $| \psi (t) \rightarrow |n+1, g \rangle$ occurs.
\end{enumerate}
\end{enumerate}
The procedure is to be repeated many times, then the simulated population should approach the analytical value obtained by Eq. (\ref{eq:pe}).

\begin{figure}
\includegraphics[width=8.5cm]{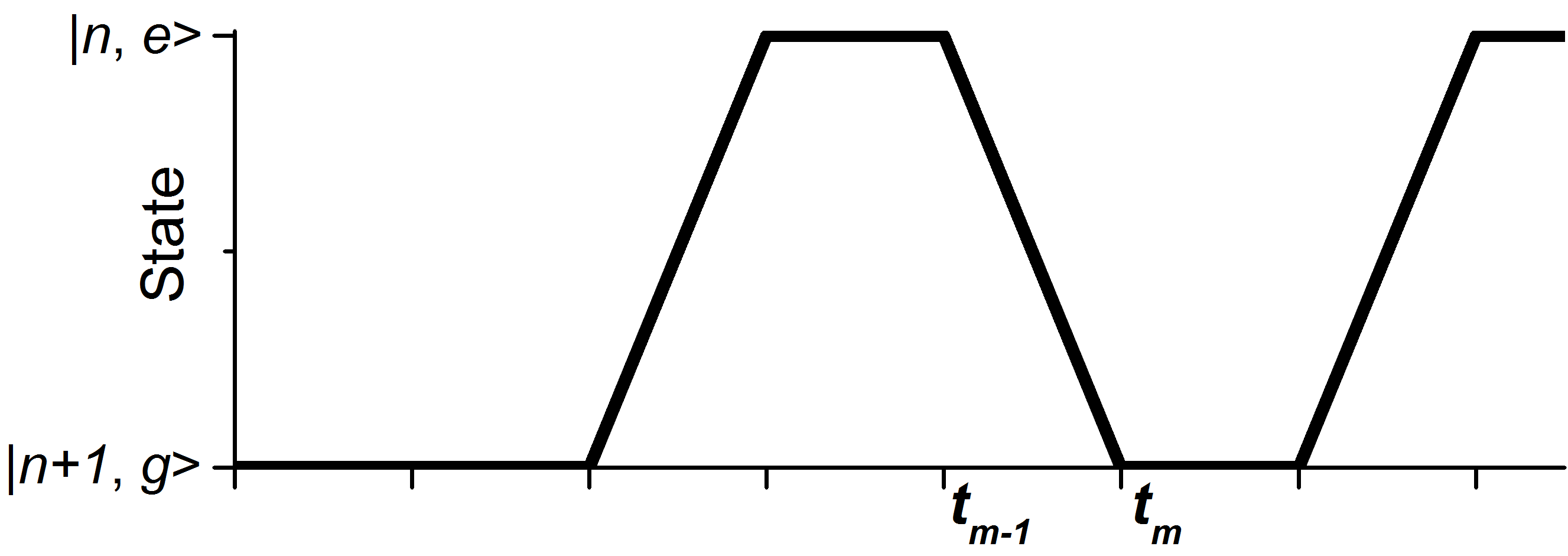}
\caption{Schematic diagram for a quantum trajectory traced by the TLA-field system. Photon emission occurs in the time interval $(t_{m-1}, t_{m}] = [m - (m-1)]\ \Delta t$.}
\label{fig:onetraj}
\end{figure}

An important benefit of the MCWF approach, unlike conventional treatments~\cite{trprob}, is that it provides a means to connect the atomic evolution with photon absorption and emission {\em dynamics}, by unraveling macroscopic observations in terms of the dynamics of individual quantum trajectories. For example, the photon emission probability in the time interval $(t_{m-1}, t_{m}]$ for a quantum trajectory is given by the joint probability
\begin{equation}
{\cal P}_{\rm emission}(t_{m}) \Delta t = P_{e}(t_{m-1}) \cdot  P_{g}(t_{m}) \Delta t .
\label{eq:probem}
\end{equation}
 A typical quantum trajectory that gives rise to a photon emission in the time interval $(t_{m-1}, t_{m}]$ is depicted schematically in Fig.~\ref{fig:onetraj}.

\subsection{MCWF Simulation Results}
We consider three initial field states to study the dependence of the atom-field dynamics on the field statistics--number, coherent, and thermal states. The latter two fields have the following photon number distributions:
\begin{eqnarray}
&p_{\rm coherent} = p_{n}(0) = \frac{{\bar n}^{n}}{n!}e^{-{\bar n} }  ,  \\
&p_{\rm thermal} = p_{n}(0) = \frac{1}{{\bar n} + 1}   \left( \frac{{\bar n}}{{\bar n} + 1} \right)^{n}. \nonumber
\label{eq:fields}
\end{eqnarray}

 In these simulations we chose a moderate value for the photon number $n=15$ for the number state and the average photon number ${\bar n}=15$ for other field states. The Rabi cycle (for the coherent and the thermal fields, the average Rabi cycle, $\Omega_{\bar n} t = 2 \pi$) was divided into $N= 10^{4}$ segments, and the vacuum Rabi frequency $\Omega_{0}$ was adjusted accordingly. The infinite sum appearing in Eq.~(\ref{eq:pe}) was truncated at the highest photon number $n_{\rm max}$ for which $p_{n_{\rm max}}(0)$ is less than $10^{-3}$ of the peak value of either of the distributions. Thus, for ${\bar n}=15$ the sum was truncated with 32 and 108 terms in the summation for the coherent and the thermal states, respectively.

\begin{figure}
\includegraphics[width=8.5cm]{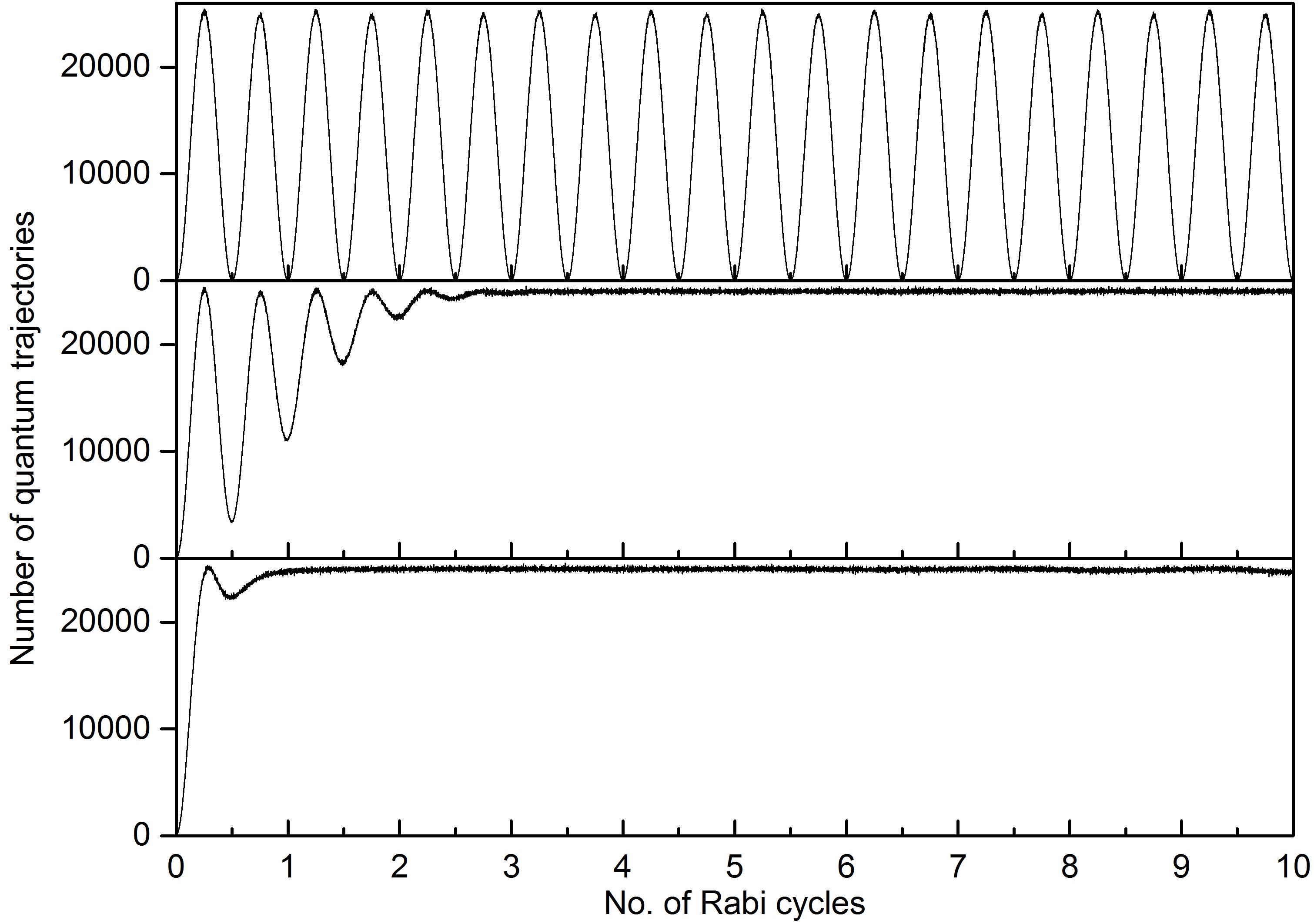}
\caption{Simulated photoemission rates as a function of time in units of Rabi cycle. From top to bottom: number, coherent, and thermal initial states. $10^{5}$ quantum trajectories are calculated for each simulation.}
\label{fig:emissions}
\end{figure}

The simulated emission rates from Eq.~(\ref{eq:probem}) are given in Fig.~\ref{fig:emissions}. For the simulation $10^{5}$ trajectory calculations are performed for the duration of 10 Rabi cycles for each field state. We can immediately see the effects of the cavity on the emission dynamics as compared to free space. The number state shows two sinusoidal oscillations per Rabi cycle with an amplitude corresponding to 1/4 of the total number of trajectories used. This oscillatory behavior is understandable, because at about $t=0$ the atom does not have enough interaction time to cause emission, despite all the atoms are initially prepared to occupy the upper state. Also, the emission rate is the lowest at about $1/2$ Rabi cycle, because at this time almost all the atoms are at the lower state. At $1/4$ and $3/4$ Rabi cycle, the atom has a $50\%$ probability to be in either the upper or the lower state, so the transition rate is the maximum there~\cite{jkps}. In the case of the coherent field the oscillation is quenched at about three Rabi cycles, and after that we see a collapse of the photon emission rate similar to the case of atomic population. The collapse time arrives much sooner for the thermal photons. A remarkable fact is that the emission rates collapse to 1/4, the ceiling value, rather than to the middle region of the `oscillations' as in the atomic population dynamics.

In order to understand this behavior we note that $t_{m-1} \rightarrow t_{m}$ as $\Delta t \rightarrow 0$, so we arrive at an asymptotic expression for the emission rate:
 \begin{align}
 \label{eq:analem}
    {\cal P}_{\rm emission} (t) &= \frac{1}{2} \left[ 1 + \sum_{n=0}^{\infty} p_{n}(0)\, \cos \, \Omega_{n} t \right] \nonumber \\
     & \times \frac{1}{2} \left[ 1 - \sum_{n=0}^{\infty} p_{n}(0)\, \cos, \ \Omega_{n} t \right].
\end{align}

\begin{figure}
\includegraphics[width=8.5cm]{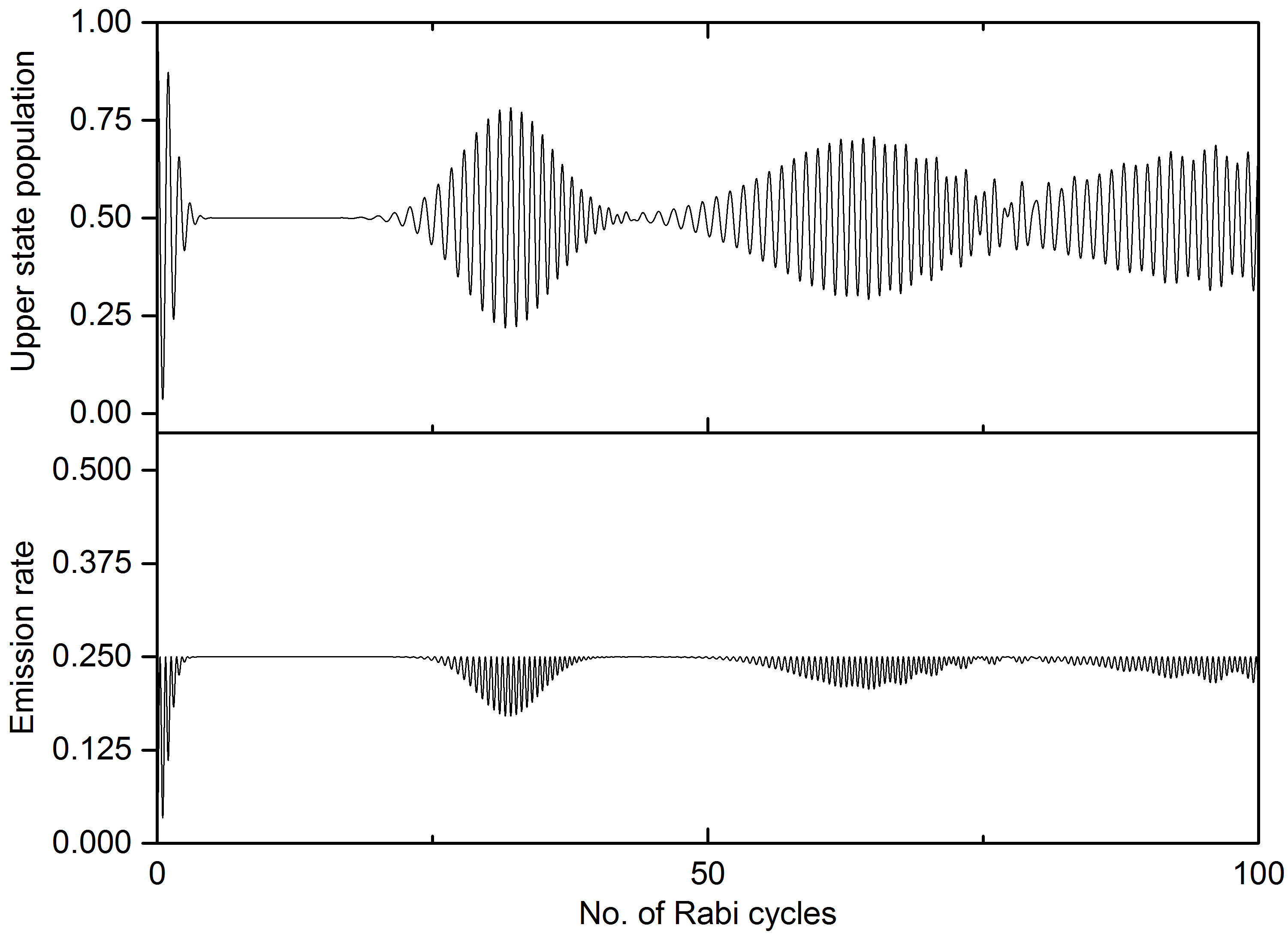}
\caption{Long-time behavior of the atomic population (top) and the photon emission (bottom) dynamics for the field initially in a coherent state, calculated using the analytic expressions, Eqs.~(\ref{eq:pe}) and~(\ref{eq:analem}). Note the difference in vertical scales.}
\label{fig:cslong}
\end{figure}

\begin{figure}
\includegraphics[width=8.5cm]{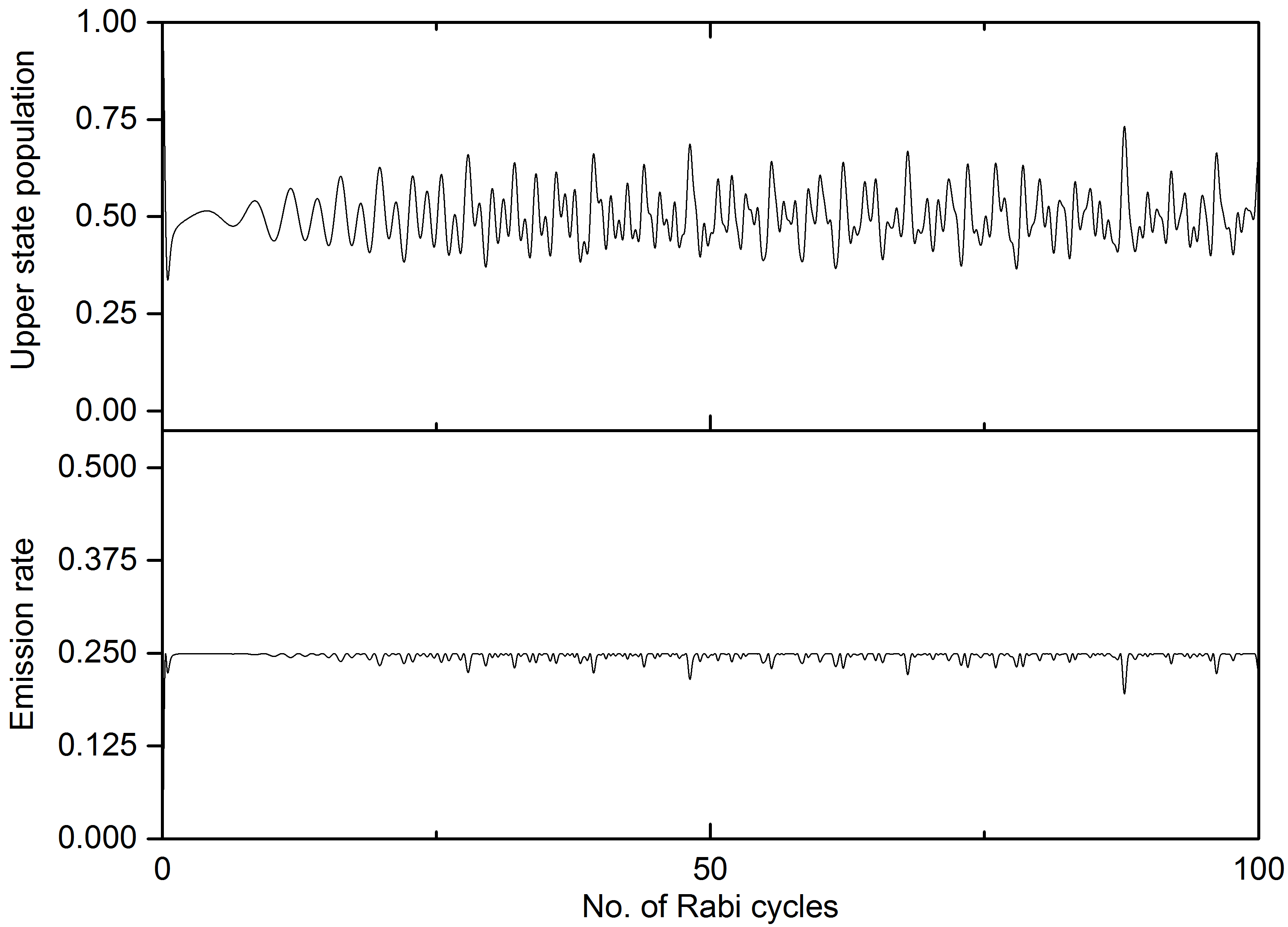}
\caption{Same as in Fig.~\ref{fig:cslong} for the field initially in a thermal state. }
\label{fig:thlong}
\end{figure}

With the analytic expression, Eq.~\eqref{eq:analem}, we can afford to investigate the long-time behavior of photon emission. Equation~(\ref{eq:analem}) was numerically evaluated with the highest $n$ values in the sum determined above, keeping all other parameters the same. In Figs.~\ref{fig:cslong} and~\ref{fig:thlong} the atomic population and the photon emission dynamics for the coherent and the thermal fields are compared up to 100 Rabi cycles. The results agree well with the corresponding simulations given in Fig.~\ref{fig:emissions}.

It is evident that for the coherent state the photon emission dynamics, in addition to the usual atomic population, also has a collapse and revival behavior. We also observe that the collapse and revival periods mirror those of the atomic population.  However, the photon emission dynamics has more rapid oscillations and spends in the collapsed state longer as compared to the atomic population dynamics. These observations may be attributed to the fact the oscillation frequency in Eq.~\eqref{eq:analem} is effectively twice bigger than that in Eq.~\eqref{eq:pe}. The ceiling value of the photon emission rate is still 1/4, which is essentially the average emission rate, $\langle \cos^{2} \, \Omega_{n}t/2  \rangle \cdot \langle \sin^{2}\, \Omega_{n}t/2 \rangle $.
For the thermal field the atomic population goes into a chaotic dynamics after a brief irregular oscillation, while the photon emission rate quickly reaches the uniform ceiling value till about 10 Rabi cycles and then goes into a chaotic regime without any sign of revival. Nonetheless, we note that the fluctuation of the photon emission rate is much smaller than that of the atomic dynamics.

\section{CONCLUSIONS}

In this paper we considered the dynamics of a coupled TLA-field system in a cavity, using the JCM from the MCWF viewpoint. The effects of the statistics of the initial light field on the dynamics of the atomic population and the photon emission were investigated. The simulation regained the familiar behavior for the atomic dynamics, for example, the collapse and revival behavior for the coherent state. More importantly, this approach enabled us to extract information on the photon emission dynamics as well, by unraveling the ensemble dynamics in terms of individual quantum trajectories. With the quantum trajectory approach, in addition to the simulations, we also obtained an analytical expression for the photon emission process that is useful for investigating long-time behavior, which is very expensive to carry out by means of simulations alone. For the coherent field in a cavity, we found that the photon emission dynamics also exhibits a series of collapses and revivals that mirror the behavior of the atomic dynamics. However, the oscillations and the collapses are not about the halfway between the maximum and minimum as in the case of atomic population, but has a ceiling value of 1/4. Photon emission for the thermal field also collapses to the same ceiling value, but shows no revival behavior. For both fields, the fluctuation in photon emission rate is seen to be much less than that in the atomic population dynamics.

The collapse region is of interest, since it may be useful for obtaining `quiet' light, by opening the cavity during this period. With the quantum trajectory approach, it is also possible to delineate the effects of initial fields on the emitted photon counting statistics in a cavity. Work along these lines are in progress.

\begin{acknowledgments}
This work was supported by Sun Moon University Research Grant of 2014.
\end{acknowledgments}


\begin{references}

\bibitem{spin_echo} E. Hahn, Phys. Rev. {\bf 80}, 580 (1950).
\bibitem{photon_echo} N. A. Kurnit, I. D. Abella, S. R. Hartmann, Phys. Rev. Lett. {\bf 13}, 567 (1964); N. A. Kurnit, I. D. Abella, S. R. Hartmann, Phys. Rev. {\bf 141}, 391 (1966).
\bibitem{vib_echo} M. D. Fayer, in {\em Ultrafast infrared vibrational spectroscopy} edited by M. D. Fayer (CRC Press, Boca Raton, 2013), p. 1.
\bibitem{atom_echo} C. J. Lee, Phys. Rev. A {\bf 53}, 4238 (1996).
\bibitem{tppi} C. J. Lee, Phys. Rev. A {\bf 58}, 3342 (1998).
\bibitem{decoherence} See, for example, M. A. Nielsen and I. L. Chuang. {\em Quantum computation and quantum information} (Cambridge University Press, New York, 2011); J.-M. Raimond and S. Haroche, in {\em Quantum Decoherence: Poincar\'e Seminar 2005}, B. Duplantier, J.-M. Raimond, and V. Rivasseau Eds. (Birkhäuser Verlag,  Basel, 2007),  p. 33.

\bibitem{jcm} E. T. Jaynes, F.W. Cummings, Proc. IEEE {\bf 51}, 89 (1963).
\bibitem{cum} F. W. Cummings, Phys. Rev. {\bf 140}, A1051 (1965).
\bibitem{eberly} J. H. Eberly, N. B. Narozhny, and J. J. Sanchez-Mondragon, Phys. Rev. Lett. {\bf 44}, 1323 (1980).
\bibitem{rempe} G. Rempe, H. Walther, and N. Klein, Phys. Rev. Lett. {\bf 58}, 353 (1987).
\bibitem{jopb_jcm}  See, for example, J. Phys. B {\bf 46}, Special issue on Jaynes¡©Cummings physics (2013).
\bibitem{mcwf} R. Dum, P. Zoller, and H. Ritsch, Phys. Rev. A {\bf 45}, 4879 (1992); H. J. Carmichael,  {\em An Open Systems Approach to Quantum Optics} (Springer-Verlag, New York, 1993); K. M{\o}lmer, Y. Castin, and J. Dalibard,  J. Opt. Soc. Am. B {\bf 10}, 524 (1993).
    \bibitem{louisell}  W. H. Louisell, {\em Quantum Statistical Properties of Radiation} (Wiley, New York, 1973).
\bibitem{bkcs} C. J. Lee, Bull. Korean Chem. Soc. {\bf 27}, 1186 (2006).
\bibitem{jkps}  C. J. Lee, J. Korean Phys. Soc. {\bf 60}, 766 (2012).
\bibitem{trprob} See, for example, L. Mandel and E. Wolf, {\em Optical Coherence and Quantum Optics} (Cambridge University Press, New York, 1995), Chap. 15.
\end{references}
\end{document}